\documentclass[twocolumn,showpacs,aps,prd]{revtex4}

\usepackage{graphicx}
\usepackage{dcolumn}
\usepackage{amsmath}
\usepackage{epsfig}

\input babarsym

\begin{document}

\def\fourpiz    {\ensuremath{\piz\piz\piz\piz}\xspace}
\def\psiprime   {\ensuremath{\psi^\prime}\xspace}
\def\fz#1       {\ensuremath{f_0({#1})}\xspace}
\def\Erad       {\ensuremath{E^*_\gamma}\xspace}
\def\besiii       {BESIII\xspace}
\def\stat       {_{\mathrm{stat}}}
\def\syst       {_{\mathrm{syst}}}
\def\etal       {{\em et al.}\xspace}
\def\ar         {\rightarrow}
\def\GG         {\gamma\gamma}
\def\bbt        {\bibitem}

\title{
{\boldmath Two-photon widths of the $\chi_{c0, 2}$ states and
helicity analysis for $\chi_{c2}\ar\gamma\gamma$} }
\date{\today}% It is always \today, today, but you may specify any date with \date.

\author{
\small
%\begin{small}
%\begin{center}
M.~Ablikim$^{1}$, M.~N.~Achasov$^{5}$, D.~J.~Ambrose$^{40}$,
F.~F.~An$^{1}$, Q.~An$^{41}$, Z.~H.~An$^{1}$, J.~Z.~Bai$^{1}$,
Y.~Ban$^{27}$, J.~Becker$^{2}$, N.~Berger$^{1}$
% ...
\\
\vspace{0.2cm}
(BESIII Collaboration)\\
\vspace{0.2cm} {\it
$^{1}$ Institute of High Energy Physics, Beijing 100049, P. R. China\\
$^{2}$ Bochum Ruhr-University, 44780 Bochum, Germany\\
% ...
$^{5}$ G.I. Budker Institute of Nuclear Physics SB RAS (BINP), Novosibirsk 630090, Russia\\
% ...
$^{27}$ Peking University, Beijing 100871, P. R. China\\
% ...
$^{40}$ University of Rochester, Rochester, New York 14627, USA\\
$^{41}$ University of Science and Technology of China, Hefei 230026, P. R. China\\
% ...
}
%\vspace{0.4cm}
%\end{small}
}

\newpage

\begin{abstract}
  Based on a data sample of 106 M $\psi^{\prime}$ events collected
  with the BESIII detector, the decays
  $\psi^{\prime}\ar\gamma\chi_{c0, 2}$,$\chi_{c0, 2}\ar\gamma\gamma$
  are studied to determine the two-photon widths of the $\chi_{c0, 2}$
  states. The two-photon decay branching fractions are determined to
  be ${\cal B}(\chi_{c0}\ar\gamma\gamma) = (2.24\pm 0.19\pm 0.12\pm
  0.08)\times 10^{-4}$ and ${\cal B}(\chi_{c2}\ar\gamma\gamma) =
  (3.21\pm 0.18\pm 0.17\pm 0.13)\times 10^{-4}$. From these, the
  two-photon widths are determined to be $\Gamma_{\gamma
    \gamma}(\chi_{c0}) = (2.33\pm0.20\pm0.13\pm0.17)$ keV,
  $\Gamma_{\gamma \gamma}(\chi_{c2}) = (0.63\pm0.04\pm0.04\pm0.04)$
  keV, and $\cal R$ $=\Gamma_{\gamma \gamma}(\chi_{c2})/\Gamma_{\gamma
    \gamma}(\chi_{c0})=0.271\pm 0.029\pm 0.013\pm 0.027$, where the
  uncertainties are statistical, systematic, and those from the PDG
  ${\cal B}(\psi^{\prime}\ar\gamma\chi_{c0,2})$ and
  $\Gamma(\chi_{c0,2})$ errors, respectively. The ratio of the
  two-photon widths for helicity $\lambda=0$ and helicity $\lambda=2$
  components in the decay $\chi_{c2}\ar\gamma\gamma$ is measured for
  the first time to be $f_{0/2}
  =\Gamma^{\lambda=0}_{\gamma\gamma}(\chi_{c2})/\Gamma^{\lambda=2}_{\gamma\gamma}(\chi_{c2})
  = 0.00\pm0.02\pm0.02$.
\end{abstract}

% PACS, the Physics and Astronomy Classification Scheme.

\maketitle

\section{Introduction}
Charmonium physics is in the boundary domain between perturbative and
nonperturbative quantum chromodynamics (QCD). Notably, the two-photon
decays of $P$-wave charmonia are helpful for better understanding the
nature of interquark forces and decay mechanisms~\cite{chao1996}. In
particular, the decays of $\chi_{c0,2}\ar\gamma\gamma$ offer the
closest parallel between quantum electrodynamics (QED) and QCD, being
completely analogous to the decays of the corresponding triplet states
of positronium. In the lowest order, for both positronium and
charmonium the ratio of the two-photon decays $\cal
R$$^{(0)}_{th}\equiv\frac{\Gamma(^3P_2\ar\gamma\gamma)}
{\Gamma(^3P_0\ar\gamma\gamma)}=4/15\approx0.27$~\cite{barbieri}.  Any
discrepancy from this simple lowest order prediction can arise due to
QCD radiative corrections and relativistic corrections, and the
measurement of ${\cal R}$ provides useful information on these
effects.  The decay of $\chi_{c1}\ar\GG$ is forbidden by the
Landau-Yang theorem~\cite{yang}.  Theoretical predictions on the decay
rates are obtained using a non-relativistic
approximation~\cite{nrqcd,barnes}, potential model~\cite{pmodel},
relativistic quark model~\cite{rquark1,rquark2}, nonrelativistic QCD
factorization framework~\cite{nrqcdf}, effective
Lagrangian~\cite{effect}, as well as lattice
calculations~\cite{lqcd}. The predictions for the ratio $\cal R\equiv$
$\frac{\Gamma_{\gamma\gamma}(\chi_{c2})}{\Gamma_{\gamma\gamma}(\chi_{c0})}$
cover a wide range values between 0.09 and
0.36~\cite{pmodel,rquark2}. Precision measurements of these quantities
will guide the development of QCD theory.

The two-photon decay widths of $\chi_{cJ}$ have been measured by many
experiments~\cite{pdg}. Using the reactions $\psi^\prime \ar \gamma
\chi_{cJ}$, the CLEO-c experiment reported results for
$\Gamma_{\GG}(\chi_{cJ})$ measured in the decay of $\chi_{cJ}$ into
two photons~\cite{cleo}:
\begin{eqnarray}
 \Gamma_{\GG}(\chi_{c0})=(2.36\pm 0.35\pm 0.22)~\mbox{keV},\\ \nonumber
\Gamma_{\GG}(\chi_{c2})=(0.66\pm0.07\pm0.06)~\mbox{keV},
\end{eqnarray}
with uncertainties that are dominated by the statistical errors.
BESIII has collected 106 million $\psi^\prime$ events, a
data sample that is about four times of that of CLEO-c, allowing for
more precise measurements of these quantities.

There are two independent helicity amplitudes, the helicity-two
amplitude ($\lambda=2$) and the helicity-zero ($\lambda=0$) amplitude,
that contribute to $\chi_{c2}\ar\gamma\gamma$ decay~\cite{barnes},
where $\lambda$ is the difference in the helicity values of the two
photons. The ratio of the two-photon partial widths for the two
helicity components, $f_{0/2}
=\Gamma^{\lambda=0}_{\gamma\gamma}(\chi_{c2})/\Gamma^{\lambda=2}_{\gamma\gamma}(\chi_{c2})$
in the decay $\chi_{c2}\ar\gamma\gamma$, is predicted to be about
0.5\%~\cite{barnes}; a measurement of this ratio can be used to
test the QCD prediction.

In this paper, $(1.06\pm0.04)\times 10^8$ $\psi^\prime$ events
accumulated in BESIII are used to study the process
$\psi^\prime \ar \gamma_1 \chi_{c0,2}$, $\chi_{c0,2} \ar \gamma_2
\gamma_3$ and measure the two-photon decay widths,
$\Gamma_{\gamma\gamma}(\chi_{c0})$ and
$\Gamma_{\gamma\gamma}(\chi_{c2})$. We also determine the ratio $\cal
R$, where many of the systematic uncertainties cancel in the ratio of
the two simultaneous measurements. The ratio of the helicity-zero
component relative to helicity-two component, $f_{0/2}$, is also
reported for the first time.

\section{The \besiii Experiment and Data Set}
\label{sec:detector}

This analysis is based on a 156.4 pb$^{-1}$ of $\psi^\prime$ data
corresponding to  $(1.06\pm0.04)\times 10^8$ $\psi^\prime$
events~\cite{number} collected with the BESIII
detector~\cite{besnim} operating at the BEPCII
Collider~\cite{bepcii}. In addition, an off-resonance sample of 44.1
pb$^{-1}$ taken at $\sqrt{s}= 3.65$ GeV is used for the study of
continuum backgrounds.

BEPCII/BESIII~\cite{besnim} is a major upgrade of the BESII
experiment at the BEPC accelerator~\cite{besii} for studies of
hadron spectroscopy and $\tau$-charm physics~\cite{besbook}. The
design peak luminosity of the double-ring $e^+e^-$ collider, BEPCII,
is $10^{33}$ cm$^{-2}$ s$^{-1}$ at a beam current of 0.93 A. The
BESIII detector has a geometrical acceptance of 93\% of $4\pi$ and
consists of four main components: (1) a small-celled, helium-based
main draft chamber (MDC) with 43 layers. The average single wire
resolution is 135 $\mu$m, and the momentum resolution for 1 GeV/$c$
charged particles in a 1 T magnetic field is 0.5\%; (2) an
electromagnetic calorimeter (EMC) made of 6240 CsI (Tl) crystals
arranged in a cylindrical shape (barrel) plus two end-caps. For 1.0
GeV photons, the energy resolution is 2.5\% in the barrel and 5\% in
the end-caps, and the position resolution is 6 mm in the barrel and
9 mm in the end-caps; (3) a time-of-flight system (TOF) for particle
identification composed of a barrel part made of two layers with 88
pieces of 5 cm thick, 2.4 m long plastic scintillators in each
layer, and two end-caps with 96 fan-shaped, 5 cm thick, plastic
scintillators in each end-cap. The time resolution is 80 ps in the
barrel, and 110 ps in the end-caps, corresponding to a 2$\sigma$
K/$\pi$ separation for momenta up to about 1.0 GeV/$c$; (4) a muon
chamber system made of 1000 m$^2$ of resistive plate chambers
arranged in 9 layers in the barrel and 8 layers in the end-caps and
incorporated in the return iron of the super-conducting magnet. The
position resolution is about 2 cm.

The optimization of the event selection and the estimation of
physics backgrounds are performed using Monte Carlo (MC) simulated
data samples. The {\sc geant4}-based simulation software
BOOST~\cite{geant4} includes the geometric and material description
of the BESIII detectors, the detector response and digitization
models, as well as the tracking of the detector running conditions
and performance. The production of the $\psi^\prime$ resonance is
simulated by the Monte Carlo event generator {\sc kkmc}~\cite{kkmc};
the known decay modes are generated by {\sc evtgen}~\cite{evtgen}
with branching ratios set at PDG~\cite{pdg} world average values,
and by {\sc lundcharm}~\cite{lund} for the remaining unknown decays.
The analysis is performed in the framework of the BESIII offline
software system~\cite{soft} which takes care of the detector
calibration, event reconstruction and data storage.

\section{Data Analysis}
\label{sec:selection}

Electromagnetic showers are reconstructed from clusters of energy
deposits in the EMC crystals. The energy deposited in nearby TOF
counters is included to improve the reconstruction efficiency and
energy resolution. Showers identified as photon candidates are
required to satisfy fiducial and shower-quality criteria. A photon
candidate is a shower detected in the EMC with a total energy deposit
greater than 25 MeV and with an angle $\theta$ with respect to the
$e^+$ beam direction in the range $|\cos\theta| < 0.75$. This
requirement is used to suppress continuum background $e^+e^- \ar
\gamma \gamma (\gamma)$, where the two energetic photons are mostly
distributed in the forward and backward regions.  We restrict the
analysis to events that have no detected charged particles. The
average event vertex of each run is assumed as the origin for the
selected candidates.  For $\psi^\prime \ar \gamma_1 \chi_{c0,2}$,
$\chi_{c0,2} \ar \gamma_2 \gamma_3$ analysis, events are required to
have three photon candidates, among which the smallest energy photon
is selected as the radiated photon $\gamma_1$ and the second-largest
and the largest energy photons are defined as $\gamma_2\gamma_3$ from
$\chi_{c0,2}$ decays.
%In this way, the miss-combinatorial rate is less than 0.03(??)\%
%according to the signal MC simulations.
An energy-momentum conservation constraint 4C-fit is performed, and
events with $\chi^{2}\leq 80$ are retained in the final selection.
The energy spectrum of the radiated photons is shown in
Fig.~\ref{fig:radgamma}, where enhancements due to the $\chi_{c0}$
and $\chi_{c2}$ over substantial backgrounds are clearly observed.

To determine signal efficiencies 100K signal MC event samples are
generated for the $\chi_{c0}$ and the $\chi_{c2}$, with PDG values for
the masses and widths~\cite{pdg}. The radiative transition
$\psi(2S)\ar\gamma_{1}\chi_{c0}$ is generated using a
$(1+\mbox{cos}^2\theta)$ distribution, where $\theta$ is the radiative
photon angle relative to the positron beam direction, in accordance
with expectations for pure E1 transitions. The $\chi_{c0}\ar \gamma_2
\gamma_3$ decays are generated using a uniform angular distribution.
Although the radiative transition $\psi(2S)\ar\gamma_{1}\chi_{c2}$ is
dominantly pure E1~\cite{e11,e12}, there is some recent experimental
evidence that the decay has contributions from higher-order
multipoles~\cite{liuzq}. The full angular amplitudes for
$\psi^\prime\ar \gamma_1 \chi_{c2}$ are discussed in association with
Eq.~(\ref{eq:angular}) in Section~\ref{sec:helicity}. Furthermore, the
$\gamma_2\gamma_3$ photons in the decay $\chi_{c2}\ar
\gamma_2\gamma_3$ are expected to be mostly in a pure helicity-two
state; the ratio of the partial two-photon widths for the
helicity-zero and helicity-two amplitudes is predicted to be less than
0.5\%~\cite{barnes}. Thus the signal MC for the decay $\psi^\prime\ar
\gamma_1 \chi_{c2}$, $\chi_{c2} \ar \gamma_2\gamma_3$ is generated
with $\gamma_2 \gamma_3$ in a helicity-two state as described in
Section~\ref{sec:helicity}.

The energy resolutions determined by the MC simulations are
$\sigma(E_{\gamma_1}) = 6.74\pm0.29$ MeV for $\chi_{c0}$ and
$\sigma(E_{\gamma_1}) = 3.91\pm0.09$ MeV for $\chi_{c2}$. The
efficiencies determined from MC simulations for the \chic{0} and
\chic{2} are $\epsilon( \chi_{c0})=(35.4 \pm 0.06)\%$ and
$\epsilon(\chi_{c2})=(38.0 \pm 0.07) \%$. The difference between
$\epsilon( \chi_{c0})$ and $\epsilon( \chi_{c2})$ is due primarily
to the different angular distributions.

The dominant non-peaking background that is apparent in the spectrum
in Fig.~\ref{fig:radgamma} is from continuum $e^+e^- \rightarrow
\gamma \gamma (\gamma)$ processes. It is determined from MC
simulations that contributions to the background due to radiative
decays to the $\eta$, $\eta^\prime$, and $3\gamma$ decays of
$\psi^\prime$ are non-peaking, spread over the full range of
$E_{\gamma_1}$, and negligible. Therefore, they do not change the
shape of the dominant continuum background. In addition we use MC
simulations to investigate possible sources of peaking
backgrounds. These are found to come from
$\chi_{c0,c2}\to\pi^{0}\pi^{0}$ and $\eta\eta$ decays and $\pi^0(\eta)
\ar \gamma \gamma$, where two of the $\gamma$s have low momentum and
are not detected or are outside of the fiducial volume of this
analysis. We generate at least 100K events of each type to determine
the efficiencies for the peaking backgrounds, and use the efficiencies
and branching fractions measured by BESIII~\cite{number} to determine
the numbers of peaking background events listed in Table
\ref{tab:peakingbg}.

\begin{table}[tb]
\begin{center}
  \caption{Expected numbers of background events peaking at the
    \chic{J} signal regions from MC simulations. The errors are the
    uncertainties from these measured branching
    fractions~\cite{number}.}
\begin{tabular} {lccc}
\hline \hline Decay modes & $n_{\chic{0}}$ & $n_{\chic{2}}$
\\\hline
$\psi^{\prime} \to \gamma \chic{0}, \chic{0} \to \piz \piz$   &  $25.4 \pm 2.2$ & $0.0 \pm 0.0$\\
$\psi^{\prime} \to \gamma \chic{0}, \chic{0} \to \eta \eta$  &  $0.4 \pm 0.1$ & $0.0 \pm 0.0$ \\
\hline
$\psi^{\prime} \to \gamma \chic{2}, \chic{2} \to \piz \piz$   &  $0.0 \pm 0.0$  & $7.7 \pm 0.7$ \\
$\psi^{\prime} \to \gamma \chic{2}, \chic{2} \to \eta \eta$   &  $0.0 \pm 0.0$  & $0.1 \pm 0.1$ \\
\hline
Sum & $25.8 \pm 2.2$ & $7.8 \pm 0.7$\\
\hline\hline
\end{tabular}
\label{tab:peakingbg}
\end{center}
\end{table}

\section{Measurement of Branching Fractions and two-photon widths}
\label{sec:fitting}

\begin{figure}[tb]
\includegraphics[width=0.8\columnwidth]{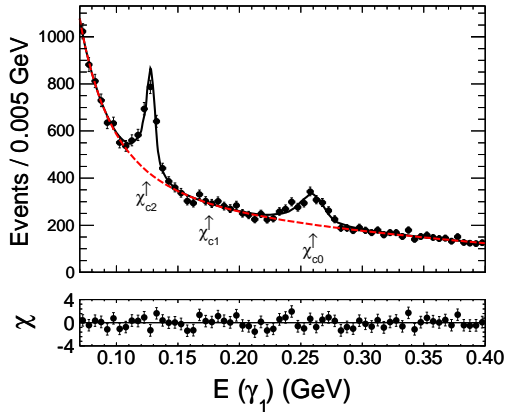}
\caption{Upper plot: the fitted $E_{\gamma_1}$ spectrum for the
$\psi^\prime$ data sample. The expected positions of $E_{\gamma_1}$
from $\chi_{c0}$, $\chi_{c1}$, $\chi_{c2}$ are indicated by arrows.
  The dashed curve shows the background line shape fixed to the
  shape in Fig.~\ref{fig:nonpeak}. Lower plot: the number of standard deviations,
  $\chi$,
  of data points from
the fitted curves.} \label{fig:radgamma}
\end{figure}

\begin{figure}[tb]
\includegraphics[width=0.8\columnwidth]{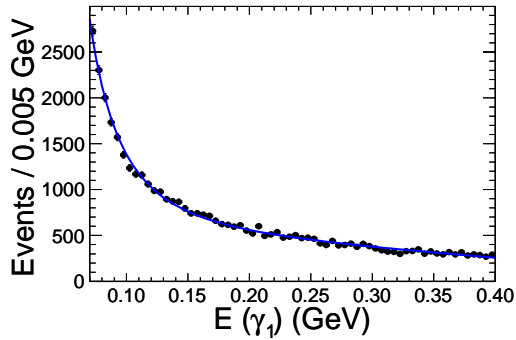}
\caption{The background $E_{\gamma_1}$ spectrum. The points are from
the off-$\psi^\prime$ data. The curve is from a fit to the
$\psi(3770)$ data.} \label{fig:nonpeak}
\end{figure}

\begin{figure}[tb]
\includegraphics[width=0.8\columnwidth]{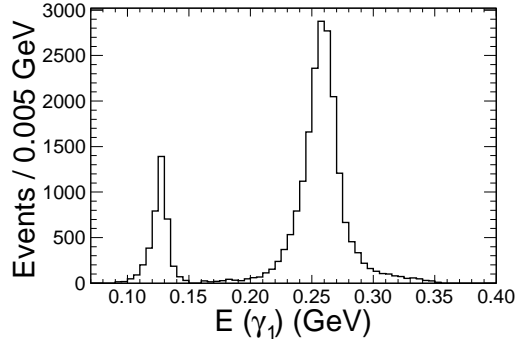}
\caption{The $E_{\gamma_1}$ spectrum for the radiative photon in the
  samples $\psi^{\prime}\ar\gamma_1 \chi_{c0,2}, \chi_{c0,2}\ar
  K^{+}K^{-}$. } \label{fig:signal}
\end{figure}

An unbinned maximum likelihood (ML) fit is done to the $E_{\gamma_1}$
spectrum as shown in Fig.~\ref{fig:radgamma}. The shape of the large
nonpeaking background in the spectrum is determined with the 44.1
pb$^{-1}$ of off-$\psi^\prime$ data taken at $\sqrt{s} = 3.65$ GeV, as
well as the $921.8$ pb$^{-1}$ of $\psi(3770)$ data taken at
$\sqrt{s}=3.773$ GeV. As is evident in Fig.~\ref{fig:nonpeak}, the
off-$\psi^\prime$ data are in good agreement with the high statistics
$\psi(3770)$ data, for which transitions to either the $\chi_{c0}$ or
$\chi_{c2}$ states are expected to be less than 8
events~\cite{pdg}. We also generate $e^+e^- \ar \gamma \gamma
(\gamma)$ MC events using the Babayaga QED event generator~\cite{yaga}
and confirm that the shapes from the 3.65 GeV and 3.773 GeV samples
are consistent with being due to the QED process. The $E_{\gamma_1}$
distribution for the $\psi(3770)$ data is fitted with the data-driven
function:
\begin{eqnarray} f_{bg}(E_{\gamma_1}) = p_{0} + p_{1}\times E_{\gamma_1} +
p_{2}\times (E_{\gamma_1})^{a}, \label{enebkgfun}
\end{eqnarray}
where $p_{0}$, $p_{1}$, $p_{2}$ and $a$ are parameters which are
obtained in a fit to the $\psi(3770)$ data in
Fig.~\ref{fig:nonpeak}. In the nominal fit to the $\psi^\prime$ data,
the background shape is fixed to Eq.~(\ref{enebkgfun}), but its
normalization is allowed to float. The shapes of the $\chi_{c0}$ and
$\chi_{c2}$ resonances used in the fit are extracted from a nearly
background-free $\psi^{\prime}\ar\gamma_1 \chi_{c0,2}, \chi_{c0,2}\ar
K^{+}K^{-}$ sample shown in Fig.~\ref{fig:signal}. The purity of the
sample is larger than 99.2\%. The shapes of the signal peaks in the
$E{\gamma_1}$ spectrum are fixed to the smoothed-histograms of the
$\psi^{\prime}\ar\gamma_1 \chi_{c0,2}, \chi_{c0,2}\ar
K^{+}K^{-}$ sample, and the yields are allowed to float. The
estimated numbers of peaking background events from $\chi_{c0,c2}\to
\pi^0\pi^0$ and $\eta\eta$ that contribute to the \chic{0} and
\chic{2} signals are 25.8 and 7.8 events, respectively, as listed
in Table~\ref{tab:peakingbg}. %The peaking background components are
%not considered in the ML fit and
They are subtracted from the fitted yields, and after this
subtraction, the signal yields are $N(\chi_{c0})=813\pm63$ and
$N(\chi_{c2})=1131 \pm 66$. The product branching fractions are
determined from the relation
\begin{eqnarray}
\mathcal{B}(\psi^\prime \to \gamma\chi_{cJ})\times
\mathcal{B}(\chic{J}\to \GG) =
\frac{N(\chi_{cJ})}{\epsilon(\chi_{cJ})\times N_{\psiprime} },
\end{eqnarray}
where $N_{\psi^\prime}$ is the total number of \psiprime in the data
sample. The measured product branching fractions are listed in
Table~\ref{tab:results}. We use the PDG average values,
\begin{eqnarray}
\mathcal{B}(\psi^\prime \to \gamma\chi_{c0}) &=& (9.68 \pm
0.31)\times 10^{-2},\\ \nonumber \Gamma(\chi_{c0}) &=& (10.4\pm 0.6)
\mbox{MeV},
\\ \nonumber
\mathcal{B}(\psi^\prime \to \gamma\chi_{c2}) &=&(8.75 \pm
0.35)\times 10^{-2}, \\ \nonumber \Gamma(\chi_{c2}) &=& (1.97\pm
0.11) \mbox{MeV},
\end{eqnarray}
to determine $\mathcal{B}(\chi_{c0,2}\to \GG)$,
$\Gamma_{\gamma\gamma}(\chi_{c0,2})$ and $\mathcal{R}$. These are
also listed in Table~\ref{tab:results}.
\begin{table}[tb]
\centering
%\begin{scriptsize}
%\begin{footnotesize}
\caption{Results of the present measurements. The first error is
statistical, second is systematic, and third is due to the PDG
values used. The common systematic errors have been removed in
determining $\mathcal{R}$. $\mathcal{B}_1\equiv
\mathcal{B}(\psi^\prime\ar \gamma\chi_{c0,2})$, $\mathcal{B}_2
\equiv \mathcal{B}(\chi_{c0,2}\ar\gamma\gamma$),
$\Gamma_{\gamma\gamma} \equiv \Gamma_{\gamma\gamma}(\chi_{c0,2}\ar\gamma\gamma)$. } %\bcl \doublerulesep 2pt
\begin{tabular}{lccc}
\hline\hline Quantity & $\chi_{c0}$ & $\chi_{c2}$ \\ \hline
$\mathcal{B}_{1}\times\mathcal{B}_{2}\times 10^{5}$ & 2.17$\pm$0.17$\pm$0.12 & 2.81$\pm$0.17$\pm$0.15 \\
$\mathcal{B}_{2} \times 10^{4}$ & 2.24$\pm$0.19$\pm$0.12$\pm$0.08 & 3.21$\pm$0.18$\pm$0.17$\pm$0.13 \\
$\Gamma_{\gamma\gamma}$ (keV) & 2.33$\pm$0.20$\pm$0.13$\pm$0.17 & 0.63$\pm$0.04$\pm$0.04$\pm$0.04 \\
$\mathcal{R}$ & \multicolumn{2}{c}{0.271$\pm$0.029$\pm$0.013$\pm$0.027} \\
 \hline\hline
\end{tabular}
\label{tab:results}
%\end{footnotesize}
%\end{scriptsize}
\end{table}

Several sources of systematic uncertainties in the measurement of
the branching fractions are considered, including: uncertainties on
the photon detection and reconstruction; the number of $\psiprime$
decays in the data sample; the kinematic fitting; the fitting
procedure and peaking background subtraction.
Table~\ref{tab:systematics} lists a summary of all sources of
systematic uncertainties. Most systematic uncertainties are
determined from comparisons of special clean, high statistics
samples with results from MC simulations.
\begin{table}[tb]
\centering \caption{Summary of systematical uncertainties of the
branching fraction measurements. Asterisks denote the systematic
sources common to both $\chi_{c0}$ and $\chi_{c2}$.}
\begin{tabular} {lccc}
\hline\hline Source of Systematic Uncertainty & $\chi_{c0}$ &
$\chi_{c2}$  \\ \hline
Number of $\psi^\prime$$^{*}$ & 4.0$\%$ & 4.0$\%$ \\
Neutral trigger efficiency$^{*}$          & 0.1\%   & 0.1\%   \\
Photon detection  $^{*}$ & 1.5$\%$ & 1.5$\%$ \\
Kinematic fit $^{*}$ & 1.0$\%$ & 1.0$\%$ \\
Resonance fitting & 3.2$\%$ & 2.9$\%$ \\
Peaking background  & 0.3\% & 0.1\% \\
Helicity 2 assumption    & -   & 0.4\%   \\
\hline Sum in quadrature & 5.5$\%$ & 5.3$\%$ \\ \hline \hline
\end{tabular}
\label{tab:systematics}
\end{table}

The number of $\psi^\prime$ events, $N_{\psi^\prime}$ , used in this
analysis is determined from the number of inclusive hadronic
$\psi^\prime$ decays following the procedure described in detail
in~\cite{number}. The result is $N_{\psi^\prime}=(1.06\pm0.04)\times
10^8$, where the error is systematic.

Three photons in the final states include a soft photon $\gamma_1$
from the radiative transition and two energetic photons
$\gamma_2\gamma_3$ from $\chi_{c0,2}$ decays. The photon detection
efficiency and its uncertainty for low energy photons are studied
using three different methods described in Ref.~\cite{guofa2011}. On
average, the efficiency difference between data and MC simulation is
less than 1\%~\cite{guofa2011}. The momenta of the two energetic
photons are more than 1.5 GeV/$c$. The systematic uncertainty due to
the reconstruction of two energetic photons is determined to be 0.25\%
per photon as described in Ref.~\cite{lei_bes3}. The total uncertainty
associated with the reconstruction of the three photons is 1.5\%.

The uncertainty due to the kinematic fit is estimated using a sample
of $e^+e^- \ar \gamma \gamma (\gamma)$, which has the same event
topology as the signal. We select the sample by using
off-$\psi^\prime$ data taken at $\sqrt{s} = 3.65$ GeV to determine the
efficiency difference between data and MC for the requirement of
$\chi^2_{4C} <80$ in the 4C-fit.  The uncertainty due to kinematic
fitting determined in this way is 1\%.

Since the signal shapes are obtained from $\psi^\prime \ar \gamma
\chi_{c0,2}$, $\chi_{c0,2}\ar K^+K^-$ events in the data, the
uncertainty due to the signal shape is negligible. The shape of the
continuum background is parameterized using the data-driven function
in Eq.~(\ref{enebkgfun}); the parameters obtained in the fitting to
off-$\psi^\prime$ data sample are fixed in the nominal fitting to
$\psi^\prime$ data. The systematic uncertainty due to the choice of
parametrization for the background shape is estimated by varying the
fitting range and the order of polynomial in our data-driven
function. We find relative changes on the $\chi_{c0}$ and $\chi_{c2}$
signal yields of 3.2\% and 2.9\%, respectively, which are taken as the
uncertainties due to the resonance fitting.

The expected numbers of peaking background events from $\chi_{c1,2}
\ar \pi^{0}\pi^{0}$ and $\chi_{c0,2} \ar \eta\eta$ decays summarized
in Table \ref{tab:peakingbg} use BESIII measurements for
$\mathcal{B}(\chi_{c1,2} \ar \pi^{0}\pi^{0}/\eta\eta)$~\cite{number}.
The uncertainties on the $\pi^0\pi^0/\eta\eta$ contributions are
estimated to be 0.3\% and 0.1\% for $\chi_{c0}$ and $\chi_{c2}$,
respectively. The systematic uncertainties due to the trigger
efficiency in these neutral channels are estimated to be $<0.1$\%,
based on cross-checks using different trigger
conditions~\cite{number,trigger}. We have assumed pure helicity-two
decay of $\chi_{c2}\ar \gamma \gamma$. In a relativistic calculation,
Barnes~\cite{barnes} predicted the helicity-zero component to be
0.5\%. In section~\ref{sec:helicity}, the ratio of the two photon
widths for the helicity-zero and helicity-two amplitudes is measured
to be $0.00\pm0.02\pm0.02$. To be conservative, we determine the
change in our $\chi_{c2}$ result when a helicity-zero component of 3\%
is included, corresponding to an upper limit at 90\% confidence level
from the measurement in this paper, to be 0.4\%, and use that as the
helicity-state-associated systematic error.

All sources of systematic errors are listed in
Table~\ref{tab:systematics}. We assume that all systematical
uncertainties are independent and add them in quadrature to obtain the
total systematical uncertainty.  For the measurements of
$\mathcal{B}(\chi_{c0,2} \ar \gamma \gamma)$, the uncertainty due to
the $\psi^\prime \ar \gamma \chi_{c0,2}$ branching fractions is kept
separate and quoted as a second systematic uncertainty.

\section{\boldmath Helicity amplitude analysis for $\chi_{c2}\ar\gamma\gamma$}
\label{sec:helicity} In $\chi_{c2}\ar\gamma\gamma$ decay, the final
state is a superposition of helicity-zero ($\lambda =0$) and
helicity-two ($\lambda=2$) components, where $\lambda$ is the
difference in the helicity values of the two photons. The formulae
for the helicity amplitudes in $\psi^{\prime}\ar\gamma\chi_{c2},
\chi_{c2}\ar\gamma\gamma$, which include higher-order multipole
amplitudes, are:
\begin{widetext}
\begin{footnotesize}
\begin{eqnarray}
 W_2(\theta_1, \theta_2, \phi_2) & = &
f_{0/2}\Bigg[\frac{3}{2}y^2(1+\cos^2\theta_{1})\sin^4\theta_2
+3x^2\sin^2\theta_1\sin^{2}2\theta_2  -
\frac{3\sqrt{2}}{2}xy\sin2\theta_1\sin^2\theta_2\sin2\theta_2\cos\phi_2\nonumber\\
& + &
\sqrt{3}x\sin2\theta_1\sin2\theta_2(3\cos^2\theta_2-1)\cos\phi_2 +
\sqrt{6}y\sin^2\theta_1\sin^2\theta_2(3\cos^2\theta_2-1)\cos2\phi_2
+(1+\cos^2\theta_1)(3\cos^2\theta_2-1)^2\bigg]_{\lambda=0}\nonumber\\
& + &
\Bigg[\frac{1}{4}y^2(1+\cos^2\theta_1)(1+6\cos^2\theta_2+\cos^4\theta_2)
+2x^2\sin^2\theta_1(1+\cos^2\theta_2)\sin^2\theta_2 +
\frac{\sqrt{2}}{4}xy\sin2\theta_1\sin2\theta_2(3+\cos^2\theta_2)\cos\phi_2\nonumber\\
& -
&\frac{\sqrt{3}}{2}x\sin2\theta_1\sin^2\theta_2\sin2\theta_2\cos\phi_2
+  \frac{\sqrt{6}}{2}y\sin^2\theta_1(1-\cos^4\theta_2)\cos2\phi_2 +
\frac{3}{2}(1+\cos^2\theta_1)\sin^4\theta_2\Bigg]_{\lambda=2}~,
\label{eq:angular}
\end{eqnarray}
\end{footnotesize}
\end{widetext}
where $x=\mathnormal{A}_1/\mathnormal{A}_0$,
$y=\mathnormal{A}_2/\mathnormal{A}_0$, and $\mathnormal{A}_{0,1,2}$
are the $\chi_{c2}$ helicity 0, 1, 2 amplitudes, respectively,
$\theta_1$ is the polar angle of the radiative photon, where the
electron beam is defined as the $z$ axis in the $e^+e^-$
center-of-mass frame, and $\theta_2$\ and $\phi_2$ are the polar angle
and azimuthal angle of one of the photons from $\chi_{c2}$ decay in
the $\chi_{c2}$ rest frame, relative to the radiative photon
direction as polar axis; $\phi_2=0$ is defined by the electron beam
direction. The factor $f_{0/2}= |F_0|^2/|F_2|^2 =
\Gamma^{\lambda=0}_{\gamma\gamma}(\chi_{c2})/\Gamma^{\lambda=2}_{\gamma\gamma}(\chi_{c2})$
is the ratio of partial two-photon widths for the helicity-zero and
helicity-two components, where $F_{0}$ ($F_{2}$) is the
helicity-zero (two) amplitude in the decay $\chi_{c2}\to
\gamma\gamma$. Further information on the formulae for the helicity
amplitudes can be found in Ref.~\cite{helicitydeduce}.

An unbinned ML fit to the angular distribution is performed to
determine $x$, $y$ and $f_{0/2}$ values. We define twelve
factors~\cite{liuzq}:
\begin{eqnarray}
a_1 = 3\sin^2\theta_1\sin^{2}2\theta_2,
\end{eqnarray}
\begin{eqnarray}
 a_2 = \frac{3}{2}(1+\cos^2\theta_{1})\sin^4\theta_2,
\end{eqnarray}
\begin{eqnarray}
 a_3 =
-\frac{3\sqrt{2}}{2}\sin2\theta_1\sin^2\theta_2\sin2\theta_2\cos\phi_2,
\end{eqnarray}
\begin{eqnarray}
a_4 =
\sqrt{3}\sin2\theta_1\sin2\theta_2(3\cos^2\theta_2-1)\cos\phi_2,
\end{eqnarray}
\begin{eqnarray}
a_5 =
\sqrt{6}\sin^2\theta_1\sin^2\theta_2(3\cos^2\theta_2-1)\cos2\phi_2,
\end{eqnarray}
\begin{eqnarray}
a_6 = (1+\cos^2\theta_1)(3\cos^2\theta_2-1)^2,
\end{eqnarray}
\begin{eqnarray}
a_7 = 2\sin^2\theta_1(1+\cos^2\theta_2)\sin^2\theta_2,
\end{eqnarray}
\begin{eqnarray}
 a_8 =
\frac{1}{4}(1+\cos^2\theta_1)(1+6\cos^2\theta_2+\cos^4\theta_2),
\end{eqnarray}
\begin{eqnarray}
a_9 =
\frac{\sqrt{2}}{4}\sin2\theta_1\sin2\theta_2(3+\cos^2\theta_2)\cos\phi_2,
\end{eqnarray}
\begin{eqnarray}
a_{10} =
-\frac{\sqrt{3}}{2}\sin2\theta_1\sin^2\theta_2\sin2\theta_2\cos\phi_2,
\end{eqnarray}
\begin{eqnarray}
a_{11} =
\frac{\sqrt{6}}{2}\sin^2\theta_1(1-\cos^4\theta_2)\cos2\phi_2,
\end{eqnarray}
\begin{eqnarray}
a_{12} = \frac{3}{2}(1+\cos^2\theta_1)\sin^4\theta_2.
\end{eqnarray}

The mean values of $a_1, \cdots, a_{12}$ can be determined with
$\psi^\prime \ar \gamma \chi_{c2}$, $\chi_{c2}\ar \gamma\gamma$ MC
events, where phase space is used for the simulation of all the
angular distributions:
\begin{eqnarray}
\bar{a}_n = \frac{\sum_{i=1}^{N}a_n(i)}{N}, n = 1, \cdots, 12
\end{eqnarray}
where $N$ is the number of events after all selections from phase
space MC samples. Since $\bar{a}_n$ is calculated with phase space
MC events after selection, it naturally accounts for the detector
acceptance effects.

The normalized probability-density function is written as:
\begin{widetext}
%\begin{footnotesize}
\begin{eqnarray}
f(x,y,f_{0/2}) = \frac{W_2(\theta_1, \theta_2, \phi_2 | x, y,
f_{0/2})} {f_{0/2}(\bar{a}_1
x^2+\bar{a}_2y^2+\bar{a}_3xy+\bar{a}_4x+\bar{a}_5y+\bar{a}_6) +
(\bar{a}_7x^2+\bar{a}_8y^2+\bar{a}_9xy+\bar{a}_{10}x+\bar{a}_{11}y
+\bar{a}_{12})}.
\end{eqnarray}
%\end{footnotesize}
\end{widetext}
A total log-likelihood function is constructed as: $
\ln\mathcal{L}=\sum_{i=1}^{n}\ln f_i(x,y,f_{0/2})$, where the sum is
over all the events in the signal region (here the signal region is
defined as $0.09<E_{\gamma_1}< 0.15$ GeV). The log-likelihood
function for the signal is given by
$\ln\mathcal{L}_{s}=\ln\mathcal{L}-\ln\mathcal{L}_{b}$, in which
$\ln\mathcal{L}_{b}$ is the normalized sum of logarithmic likelihood
values from background events and is calculated using the events in
the sidebands, which are defined in the ranges $(0.07, 0.08)$ GeV
(lower sideband) and $(0.16,0.20)$ GeV (higher sideband) in the the
$E_{\gamma_{1}}$ spectrum. By maximizing the logarithm of the
likelihood function $\ln\mathcal{L}_{s}$, the best values of $x$,
$y$ and $f_{0/2}$ are determined. Before fitting to the data, input
and output checks were done using MC samples, and the results used
to validate the fitting procedure.

\begin{figure}[tb]
\includegraphics[width=0.7\columnwidth]{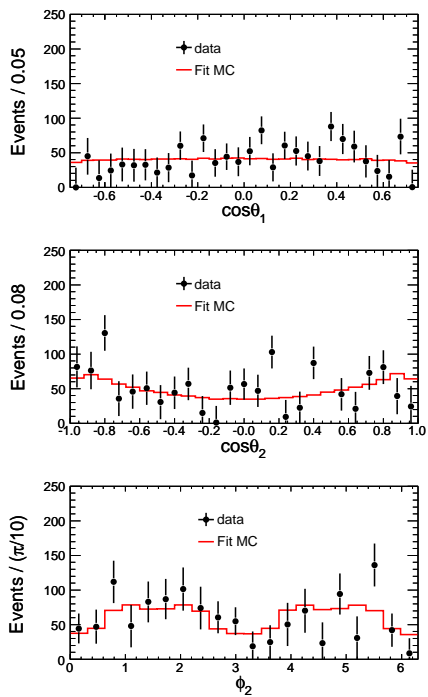}
\caption{Distributions of $\cos\theta_{1}$, $\cos\theta_{2}$ and
$\phi_{2}$ for the $\psi^\prime\to \gamma \chi_{c2}$, $\chi_{c2}\to
\gamma \gamma$, where the dots with error bars are
background-subtracted data and the histograms are the fitted
results.} \label{fig:angular}
\end{figure}

 BESIII has
determined   $x$ and $y$ to be $x=1.55\pm
0.05(\mbox{stat.})\pm0.07(\mbox{syst.})$ and
$y=2.10\pm(\mbox{stat.})0.07\pm0.05(\mbox{syst.})$~\cite{liuzq}
using the decays $\psi^\prime \ar \gamma \chi_{c2}$,
$\chi_{c2}\ar\pi^{+}\pi^{-}/K^{+}K^{-}$. Therefore, in the nominal
fit, the $x$ and $y$ parameters are fixed to the measured values, and
the remaining  parameter $f_{0/2}$ is determined to be:
\begin{eqnarray}
f_{0/2} = 0.00\pm0.02 ,
\end{eqnarray}
where the error is statistical.  Figure~\ref{fig:angular} shows the
angular distributions of background-subtracted data and the fitted
results for $\chi_{c2}\ar\gamma\gamma$ events. It is found that all
angular distributions are consistent with the fitted results within
errors. As mentioned in Section~\ref{sec:fitting}, for the
measurements of the branching fractions we use the formulae in
Eq.~(\ref{eq:angular}) to generate MC events for efficiency
determination of $\psi^\prime \ar \gamma \chi_{c2}$, $\chi_{c2}\ar
\gamma\gamma$ decay, with the $x$, $y$ and $f_{0/2}$ parameters
fixed at their measured central values ($x=1.55$, $y=2.10$ and
$f_{0/2}= 0.0$).

In order to test the reliability of the fit, we allow the $x$ and
$y$ parameters to float  in the fit, in which case the likelihood
fit to data yields
\begin{eqnarray}
x = 1.76\pm0.25 , y=2.00\pm0.23,
\end{eqnarray}
where the errors are statistical. The results are consistent with
the previous BESIII measurements of the $\psi^\prime \ar \gamma
\chi_{c2}$, $\chi_{c2}\ar\pi^{+}\pi^{-}/K^{+}K^{-}$
decays~\cite{liuzq}.

The goodness of the fit is estimated using Pearson$^{\prime}$s
$\chi^{2}$ test ~\cite{eadie}. The data and MC are divided into 6
bins of equal width  in each dimension (i.e. $\cos\theta_1$,
$\cos\theta_2$, $\phi_2$) of the three-dimension angular
distribution. The numbers of events in each cell for data and the
normalized MC sample are compared. The $\chi^2$ is defined as:
\begin{eqnarray}
\label{chisq} \chi^2  = \sum_i
\frac{(n^{DT}_i-n^{MC}_i)^2}{\sigma^2_{n^{DT}_i}},
\end{eqnarray}
where $n^{DT}_i$ ($\sigma_{n^{DT}_i}$) is the observed number (its
error) of signal events after background subtraction in the $i$th
bin from data and $n^{MC}_i$ is the expected number of events
predicted from MC in the $i$th bin using $f_{0/2}$ fixed to the
value determined in the analysis. If the number of events in a bin
is less than 5, we add the events to the adjacent bin. The result of
the $\chi^2$ test of the fitting is: $\chi^2/n.d.f = 87.3/88 =
0.99$, where $n.d.f.$ is the number of degrees of freedom. The
result shows that the fit quality is acceptable.

Since $f_{0/2}$ is a ratio, many systematic errors cancel, and
only the effects due to MC simulation of detector response, the
uncertainties on the measured $x$ and $y$ parameters, background
substraction, $\chi_{c0}$ contamination are considered here. Among
these sources of the systematic uncertainties, the MC simulation of
detector response is dominant; the others are tiny and are
neglected.

As discussed above,  the $x$ and $y$ parameters are fixed to the
measured values from Ref.~\cite{liuzq} in the ML fit to
$\chi_{c2}\rightarrow \gamma \gamma$ events in order to obtain the
ratio $f_{0/2}$. In the fit we change the $x$ and $y$ central values
by one standard deviation of the measured values~\cite{liuzq}, and
find that the effect on $f_{0/2}$ is negligible. To estimate the
uncertainty due to background subtraction, we vary the sideband
region from $(0.07, 0.08)$ GeV (lower sideband) and $(0.16,0.20)$
GeV (higher sideband) to $(0.07, 0.09)$ GeV and $(0.15,0.20)$ GeV.
After subtraction of the background based on  the sum of
recalculated logarithmic likelihood values, $\ln\mathcal{L}_{b}$, we
find that the fitted $f_{0/2}$ value is almost unchanged. From MC
simulation, 0.028\% of the $\chi_{c0}\rightarrow \gamma \gamma$
events are distributed under the $\chi_{c2}$ signal region; the
uncertainty due to $\chi_{c0}$ contamination is estimated to be
negligible.

The uncertainty due to the inconsistency between data and MC
simulation on the angular distributions for $\chi_{c2}$ events can be
tested using $\chi_{c0}$ events. Since the $\chi_{c0}$ is pure
helicity-zero, the $x$ and $y$ parameters are expected to be zero. In
$\chi_{c0} \rightarrow \gamma \gamma$ decay, the difference of
helicity values of the two photons is also expected to be zero, so
only the helicity-zero term in Eq.~(\ref{eq:angular}) remains, which
modifies Eq.~(\ref{eq:angular}) to:
\begin{widetext}
\begin{footnotesize}
\begin{eqnarray}
W_0(\theta_1, \theta_2, \phi_2) & = &
\Bigg[\frac{3}{2}y^2(1+\cos^2\theta_{1})\sin^4\theta_2
+3x^2\sin^2\theta_1\sin^{2}2\theta_2  -
\frac{3\sqrt{2}}{2}xy\sin2\theta_1\sin^2\theta_2\sin2\theta_2\cos\phi_2\nonumber\\
& + & \sqrt{3}x\sin2\theta_1\sin2\theta_2\cos\phi_2 +
\sqrt{6}y\sin^2\theta_1\sin^2\theta_2\cos2\phi_2
+(1+\cos^2\theta_1)\bigg]_{\lambda=0}\nonumber\\
& + &
f_{2/0}\Bigg[\frac{1}{4}y^2(1+\cos^2\theta_1)(1+6\cos^2\theta_2+\cos^4\theta_2)
+2x^2\sin^2\theta_1(1+\cos^2\theta_2)\sin^2\theta_2 +
\frac{\sqrt{2}}{4}xy\sin2\theta_1\sin2\theta_2(3+\cos^2\theta_2)\cos\phi_2\nonumber\\
& -
&\frac{\sqrt{3}}{2}x\sin2\theta_1\sin^2\theta_2\sin2\theta_2\cos\phi_2
+  \frac{\sqrt{6}}{2}y\sin^2\theta_1(1-\cos^4\theta_2)\cos2\phi_2 +
\frac{3}{2}(1+\cos^2\theta_1)\sin^4\theta_2\Bigg]_{\lambda=2}~,
 \label{eq:angular-0}
\end{eqnarray}
\end{footnotesize}
\end{widetext}
where the product factor $f_{0/2}$ is moved to the front factor of
the helicity-two term and renamed as $f_{2/0}$, and the
$(3\cos^2\theta_2-1)^2$ term associated with $\lambda=0$ amplitude
in Eq.~(\ref{eq:angular}) is replaced by 1, so that one can obtain
the expected angular distribution $W_0 = 1+\cos^2\theta_1$ from
Eq.~(\ref{eq:angular-0}) if the parameters $x=0$, $y=0$ and
$f_{2/0}=0$, as expected. Therefore, we fit the angular distribution
of $\chi_{c0}$ with the Eq.~(\ref{eq:angular-0}) using the same
method as in $\chi_{c2}$ decays; non-zero $x$, $y$ and $f_{2/0}$
values will indicate the inconsistency between data and MC
simulation.  The systematic error is taken as the shift from 0 plus
its error. The fitted results are $x=-0.11\pm0.09$, $y=0.13\pm0.07$
and $f_{2/0}=0.00\pm 0.02$. The correlation coefficient between $x$
and $y$ is -0.27, while it is 0.0 between $x$ ($y$) and $f_{2/0}$.
Thus we take 0.02 as the systematic error for the measurement of
$f_{0/2}$ in the fit to $\chi_{c2}$ events. Studies with  MC
simulated data samples demonstrate that a systematic error in
modeling the $\theta_1$, $\theta_2$, and $\phi_2$  efficiency
produces a shift of approximately the same size for $f_{2/0}$ in
$\chi_{c0}$ sample
 and $f_{0/2}$ in $\chi_{c2}$ sample, when the latter sample is generated with $x=1.55$, $y=2.10$ and
 $f_{0/2}=0$. Therefore, we assume the observed shift from $f_{2/0}$
for the true $\chi_{c0}$ data is an estimate of the systematic error
on the measured values of $f_{0/2}$ for the two-photon decay of
$\chi_{c2}$.
\begin{table*}[tbh]
\centering
%\begin{scriptsize}
%\begin{footnotesize}
\caption{The comparison of experimental results for the two-photon
partial widths of $\chi_{c0}$
and $\chi_{c2}$. } %\bcl \doublerulesep 2pt
\begin{tabular}{lccccc}
\hline\hline Quantity & PDG global fit results$^{a}$ & CLEO-c$^{b}$
& This measurement$^{b}$\\\hline
$\mathcal{B}_{1}\times\mathcal{B}_{2}\times 10^{5}(\chi_{c0})$$^{c}$ & $2.16\pm0.18$ & $2.17\pm0.32\pm0.10$ & $2.17\pm0.17\pm0.12$\\
$\mathcal{B}_{1}\times\mathcal{B}_{2}\times 10^{5}(\chi_{c2})$$^{c}$& $2.24\pm0.17$ & $2.68\pm0.28\pm0.15$ & $2.81\pm0.17\pm0.15$\\
$\mathcal{B}_{2} \times 10^{4}(\chi_{c0})$$^{c}$ & $2.23\pm0.17$ & $2.31\pm0.34\pm0.15$ & $2.24\pm0.19\pm0.15$\\
$\mathcal{B}_{2} \times 10^{4}(\chi_{c2})$$^{c}$ & $2.56\pm0.16$ & $3.23\pm0.34\pm0.24$ & $3.21\pm0.18\pm0.22$\\
$\Gamma_{\GG}(\chi_{c0})$(keV) &$2.32\pm0.22$&$2.36\pm0.35\pm0.22$&$2.33\pm0.20\pm0.22$ \\
$\Gamma_{\GG}(\chi_{c2})$(keV) & $0.50\pm0.05$ & $0.66\pm0.07\pm0.06$ & $0.63\pm0.04\pm0.06$ \\
${\cal R}$ & $0.22\pm0.03$ & $0.28\pm0.05\pm0.04$ & $0.27\pm0.03\pm0.03$ \\
$f_{0/2}$ & - & - & $0.00\pm0.02\pm0.02$ \\
\hline\hline
\end{tabular}
\label{finalresult}
%\end{footnotesize}
%\end{scriptsize}
\begin{flushleft}
$^{a}$ The results from the literature have been reevaluated by
using the branching fractions and total widths from the PDG global
fit.

$^{b}$ The first error is statistical. The second error is
systematic error combined in quadrature with the error in the
branching fractions and widths used.

$^{c}$ $\mathcal{B}_{1}\equiv\mathcal{B}(\psi(2S) \to \gamma
\chi_{c0,c2})$, $\mathcal{B}_{2}\equiv\mathcal{B}(\chi_{c0,c2} \to
\gamma\gamma)$,
$\Gamma_{\gamma\gamma}(\chi_{c0,c2})\equiv\Gamma_{\gamma\gamma}(\chi_{c0,c2}
\to \gamma\gamma)$.
\end{flushleft}
\end{table*}

\section{Conclusion}
\label{sec:summary}

In summary, we present measurements of  the two-photon decays of
$\chi_{c0,2}$ via the radiative decays $\psi^\prime \ar \gamma
\chi_{c0,2}$ . We find $\mathcal{B}(\chi_{c0} \to \gamma \gamma)
=(2.24\pm0.19\pm0.15)\times 10^{-4} $ and $\mathcal{B}(\chi_{c2} \to
\gamma \gamma) =(3.21\pm0.18\pm0.22)\times 10^{-4} $, which agree
with the results from the CLEO experiment~\cite{cleo}. The partial
widths $\Gamma_{\gamma\gamma}(\chi_{c0,c2})$ and the ratio
$\mathcal{R}$ of the two-photon partial widths between $\chi_{c2}$
and $\chi_{c0}$ are determined from these measurements. The precision
of our measurements is improved compared to CLEO's; the final
results are listed in Table~\ref{finalresult}.

Since theoretical unknowns cancel in the ratio $\mathcal{R}$, a
calculation including the first-order radiative corrections by
Voloshin~\cite{voloshin} predicts
$\mathcal{R}^{(1)}_{\mbox{th}}=0.116\pm0.010$. Our experimental
result, $\mathcal{R}=0.27\pm0.04$, indicates some inadequacy of the
first-order radiative corrections that have been used to make
theoretical predictions for charmonium decays.

We also perform a helicity amplitude analysis for the decay of
$\psi^\prime \to \gamma \chi_{c2}$, $\chi_{c2}\to \gamma \gamma$;
the ratio of the two-photon partial widths for the helicity-zero and
helicity-two components in the decay $\chi_{c2}\ar\gamma\gamma$ is
determined for the first time to be $f_{0/2} = 0.00\pm0.02\pm0.02$.
The helicity-zero component in the $\chi_{c2} \to \gamma \gamma$
decay is highly suppressed. This measurement is consistent with the
calculations based on a relativistic potential model~\cite{barnes},
in which the ratio is predicted to be less than 0.5\%.

\section{Acknowledgments}
%V23-11-2010
The BESIII collaboration thanks the staff of BEPCII and the
computing center for their hard efforts. One of the authors, H.~B.~Li,
 would like to thank Ted Barnes for useful discussions. This work is
supported in part by the Ministry of Science and Technology of China
under Contract No. 2009CB825200; National Natural Science Foundation
of China (NSFC) under Contracts Nos. 10625524, 10821063, 10825524,
10835001, 10935007, 11125525, 10975143; Joint Funds of the National
Natural Science Foundation of China under Contracts Nos. 11079008,
11179007; the Chinese Academy of Sciences (CAS) Large-Scale
Scientific Facility Program; CAS under Contracts Nos. KJCX2-YW-N29,
KJCX2-YW-N45; 100 Talents Program of CAS; Istituto Nazionale di
Fisica Nucleare, Italy; U. S. Department of Energy under Contracts
Nos. DE-FG02-04ER41291, DE-FG02-91ER40682, DE-FG02-94ER40823; U.S.
National Science Foundation; University of Groningen (RuG) and the
Helmholtzzentrum fuer Schwerionenforschung GmbH (GSI), Darmstadt;
WCU Program of National Research Foundation of Korea under Contract
No. R32-2008-000-10155-0.

\end{document}